\documentclass[11pt,a4paper]{article}
\usepackage{jinstpub}
\usepackage{float}
\usepackage{url}
\usepackage{lineno,hyperref}
\usepackage{caption}
\usepackage{subcaption}
\usepackage[utf8]{inputenc} 

\modulolinenumbers[5]

\begin{document}
	\title{Observation of unusual slow components in electroluminescence signal of two-phase argon detector}

	\author[a,b]{A.~Bondar,}
	\author[a,b]{E.~Borisova,}
	\author[a,b]{A.~Buzulutskov,}
	\author[a,b,1]{E.~Frolov,
	\note{Corresponding author.}}
	\author[a,b]{V.~Oleynikov,}
	\author[a,b]{A.~Sokolov}

	\affiliation[a]{Budker Institute of Nuclear Physics SB RAS, Lavrentiev avenue 11, 630090 Novosibirsk, Russia}
	\affiliation[b]{Novosibirsk State University, Pirogova st. 2, Novosibirsk 630090, Russia}

	\emailAdd{A.F.Buzulutskov@inp.nsk.su (A. Buzulutskov)}
	\emailAdd{geffdroid@gmail.com (E. Frolov)}

	\abstract{Proportional electroluminescence (EL) in noble gases is used in two-phase detectors for dark matter search to record ionization signals in the gas phase induced by particle scattering in the liquid phase (S2 signals). In this work, the EL pulse-shapes in a two-phase argon detector have for the first time been studied systematically in a wide range of reduced electric field, varying from 3 to 9~Td. The pulse-shapes were studied at different readout configurations and spectral ranges: using cryogenic PMTs and SiPMs, with and without a wavelength shifter (WLS), in the VUV and visible range. We observed the fast component and two unusual slow components, with  time constants of about 5~$\mu$s and 40~$\mu$s. The unusual characteristic property of slow components was that their contribution and time constants increased with electric field.}


\maketitle 

	\section{Introduction}
	Two-phase detectors with electroluminescence (EL) gap based on Ar or Xe are relevant to experiments for direct search of dark matter particles \cite{Chepel:survey, DarkSide20k18, Aprile18:XeTPC1T}. In these detectors, primary scintillation (S1 signal) and primary ionization (S2 signal) caused by interaction of particle with liquid target are detected. The S2 signal is detected through the effect of proportional electroluminescence (or secondary scintillation) \cite{Oliveira2011:EL_simulation, Buzulutskov17:ELReview} in electroluminescence (EL) gap located directly above the liquid-gas interface.
An understanding of S2 pulse-shapes is essential for correct analysis and interpretation of the data. This is especially crucial in low mass WIMP search using ``S2 only'' analysis where only the S2 signal is taken into consideration \cite{Agnes2018_DS50_S2_only}. Overall, there are several motivations for in-depth study of the S2 pulse-shape in a wide range of electric fields:
	\begin{itemize}
	  \item Correct integration times should be used in order to determine the overall S2 amplitude.
	  \item The smearing of the S2 pulse-shape is mostly caused by diffusion of electron cloud in the liquid. Utilizing this, z-coordinate of interaction point can be measured from the S2 pulse-shape \cite{Agnes2018}.
	  \item The width of the S2 pulse-shape corresponds to drift time through the EL gap at a given x-y position. This can be used to measure the EL gap thickness in situ and monitor the liquid level and potential sagging of electrodes.
	  \item The S2 pulse-shape analysis may be additional tool for the study of the EL mechanisms. For example, it is known that scintillation in gaseous Ar proceeds through Ar$_{2}^{*}(^{1,3}{\Sigma}_{u}^{+})$ singlet and triplet excimer states resulting in 128~nm VUV emission with fast (4.2~ns) and slow (3.1~$\mu$s) components respectively \cite{Buzulutskov17:ELReview}. The triplet time constant and the ratio of the singlet-to-triplet state population can be extracted from the fit of the S2 pulse-shapes \cite{Agnes2018}. In addition, an alternative electroluminescence mechanism, namely that of neutral bremsstrahlung \cite{NBrS18, NBrS20, Tanaka2020:NBrS_Ar, Takeda2020:NBrS_Ar}, significantly contributes to the fast component at lower electric fields.
	\end{itemize}
	More detailed overview on the S2 pulse-shape problem in Xe and Ar will be given in the next section~\ref{PulseShapeOverview}.

	Several versions of two-phase detectors were developed in our laboratory for the study of the EL mechanism in pure argon \cite{NBrS18, NBrS20} and argon doped with nitrogen \cite{ArN2CRAD17,Buzulutskov17:ELReview}, for SiPM-matrix readout of two-phase detectors, directly \cite{SiPMMatrix19} or using combined THGEM/SiPM-matrix multipliers \cite{SiPMMatrix19,CRADRev12,SiPMMatrix13}, and for the measurements of ionization yields in liquid argon \cite{IonYield17}. 
	
	In this work, two configurations of the detector with PMT and SiPM-matrix readout were used: with tetraphenyl-butadiene (TPB) wavelength shifter (WLS) films deposited in front of PMTs and without ones. This allowed us to systematically study electroluminescence (S2) pulse shape in Ar for devices with different spectral sensitivity in a wide range of reduced electric fields ($E/N$), from 3.4 to 8.5~Td (1~Td = $10^{-17}$~V~cm$^2$ corresponding to 0.87~kV/cm in gaseous argon at 87.3~K). In particular, in addition to the usual triplet excimer 3.1~$\mu$s component \cite{Buzulutskov17:ELReview}, two unusual slow components were observed, with time constants of about 5~$\mu$s and 40~$\mu$s, further referred to as slow and long component respectively. In this work, their properties as a function of the electric field are studied and their possible nature is discussed.
	
	\section{Slow components in S2 signals of two-phase detectors}\label{PulseShapeOverview}

	In two-phase detectors, the primary ionization produced by a scattered particle is first drifted in the liquid phase and then emitted into the gas phase where it produces electroluminescence (S2) signal during drifting in the EL gap. Thus, the pulse-shape of the S2 signal is determined by the drift time over the EL gap and the shape of the ionization electron cluster reaching the gas phase. Usually, events of interest in two-phase detectors are those of single scattering (i.e. localized events). In that case, if the electron emission through the liquid-gas interface occurs without delays, all the  electroluminescence processes are sufficiently fast (much faster then the electron drift time through the EL gap) and the light collection efficiency is uniform across the drift path, then the S2 pulse-shape is described by a simple rectangular function smeared according to the distribution of the electrons in the cluster. The electron cluster size is defined by diffusion during the drift through the liquid. In this case, the electron distribution has a Gaussian shape and the S2 pulse-shape is a convolution of rectangular function with said Gaussian. This is a typical form of so-called fast component of electroluminescence signal.

	In practice, however, some or all of the three conditions are unsatisfied. Among them, the light collection uniformity depends on the detector geometry only and its non-uniformity can be accounted for by MC simulations. 
	
	The electron emission from the liquid is known to have prompt (fast) and delayed (slow) components in both Ar and Xe due to the presence of potential barrier at the interface \cite{Bolozdynya1999, Gushchin1982, Aprile2004:XeTPC, Borghesani1990, Bondar2009:Emission}. Delayed component contribution and time constant decrease with the increase of the electric field. The latter was measured for Ar \cite{Borghesani1990, Bondar2009:Emission} and it decreased according to the thermionic emission model \cite{Borghesani1990} as $\sim 1/E$: from about 900~$\mu$s at 0.1~kV/cm to 5~$\mu$s at 1.6~kV/cm.
	As a result, delayed component could be present in EL signal at low electric fields. However, most of the two-phase detectors operate at high electric fields where 100\% electron emission efficiency is achieved in which case the contribution of this delayed component to the pulse shape becomes negligible.		

	Regarding EL processes, photon emission occurs via excimer formation in both Ar and Xe at 128 and 175~nm respectively. Both gases have singlet $^{1}{\Sigma}_{u}^{+}$ and triplet $^{3}{\Sigma}_{u}^{+}$ excimer states which decay to VUV with time constants of 4.2~ns and $\sim$3.1~$\mu$s respectively for Ar and 4.5~ns and 100~ns for Xe \cite{Buzulutskov17:ELReview, Keto1974:ArXeExcimerEmission, Policarpo1981:ArEximerEmission, Suzuki1982:ArXeExcimerEmission, Moutard88:XeExcimerDecay, Morikaw89:ArXeKrExcimerEL}. Since only Ar excimers have appreciably large decay time, only S2 signal in Ar should have distinct slow component, in microsecond scale, observable in two-phase detectors \cite{Agnes2018}.

	On the other hand, some strange slow components were observed in two-phase Xe detectors on microsecond and millisecond scales. In particular, \cite{Aprile2014:XeDelayedEmission} showed that delayed electrons were correlated to S2 and large S1 signals and were produced via photoionization of impurities or detector surfaces. Also, \cite{Akimov2016:XeDelayedEmission} argued that electrons were trapped in the liquid under the surface and due to the tilt of the detector they drifted to its edge where they were emitted in the gas producing so-called delayed emission. Finally, \cite{Sorensen2018:XeDelayedEmission} observed slow ($\sim$40~$\mu$s) and long ($\sim$2~ms) components with unusual character, their decay time increasing with the electric field. The contribution of long component was also increasing with the field. The authors believe their observations can be explained by photoionization of impurities and that long component also could be explained if there were some higher-level Xe states with millisecond lifetimes.

	In contrast to Xe, there have been no systematic study of S2 slow components in Ar at different electric fields. This has been done in this work.
	
	\section{Experimental setup}\label{SetupSection}

\begin{figure}[t]
	\centering
	\includegraphics[width=0.7\linewidth]{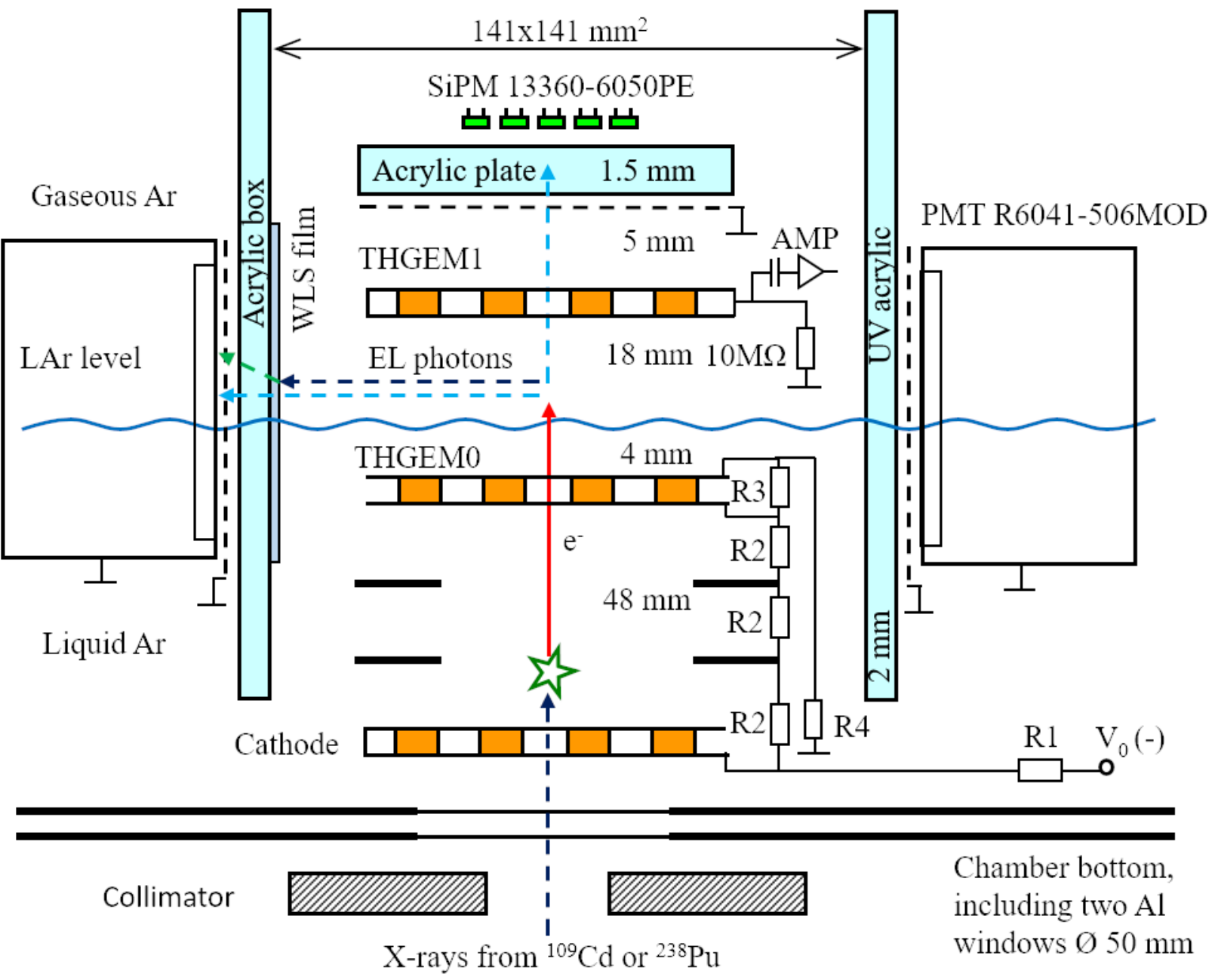}
	\caption{Generalized version of the two-phase detector with EL gap used in \cite{NBrS18, NBrS20, ArN2CRAD17,SiPMMatrix19,IonYield17} (not to scale). AMP is a charge sensitive amplifier used for direct charge measurement with Mo pulsed X-ray tube. Two detector configurations were used in this work: first, with  WLS (TPB in polystyrene matrix) deposited in front of 3 (out of 4 total) photomultipliers, and second, without WLS in the setup at all. Relevant transmittances, QE, PDE and WLS emission spectra are shown in Fig.~\ref{fig:spectra}.}
	\vspace{-10pt}
	\label{fig:setup_scheme}
	\end{figure}

	The generalized version of the two-phase detector with EL gap used in the current studies \cite{NBrS18, NBrS20,SiPMMatrix19} is shown in Fig.~\ref{fig:setup_scheme}. To form the drift, electron emission and electroluminescence regions,  THGEM (Thick Gas Electron Multiplier) electrodes were used instead of more conventional wire grids, providing the advantage of electrode rigidity  that allowed to avoid the problem of wire grid sagging. The detector included a cathode electrode, field-shaping electrodes and THGEM0 (interface THGEM), immersed in a 55~mm thick liquid Ar layer. These elements were biased through a resistive high-voltage divider placed within the liquid, forming a 48 mm long drift region in liquid Ar. A 4~mm thick liquid Ar layer above the THGEM0 acted as an electron emission region.

	\begin{figure}[t]
	\centering
	\includegraphics[width=0.7\linewidth]{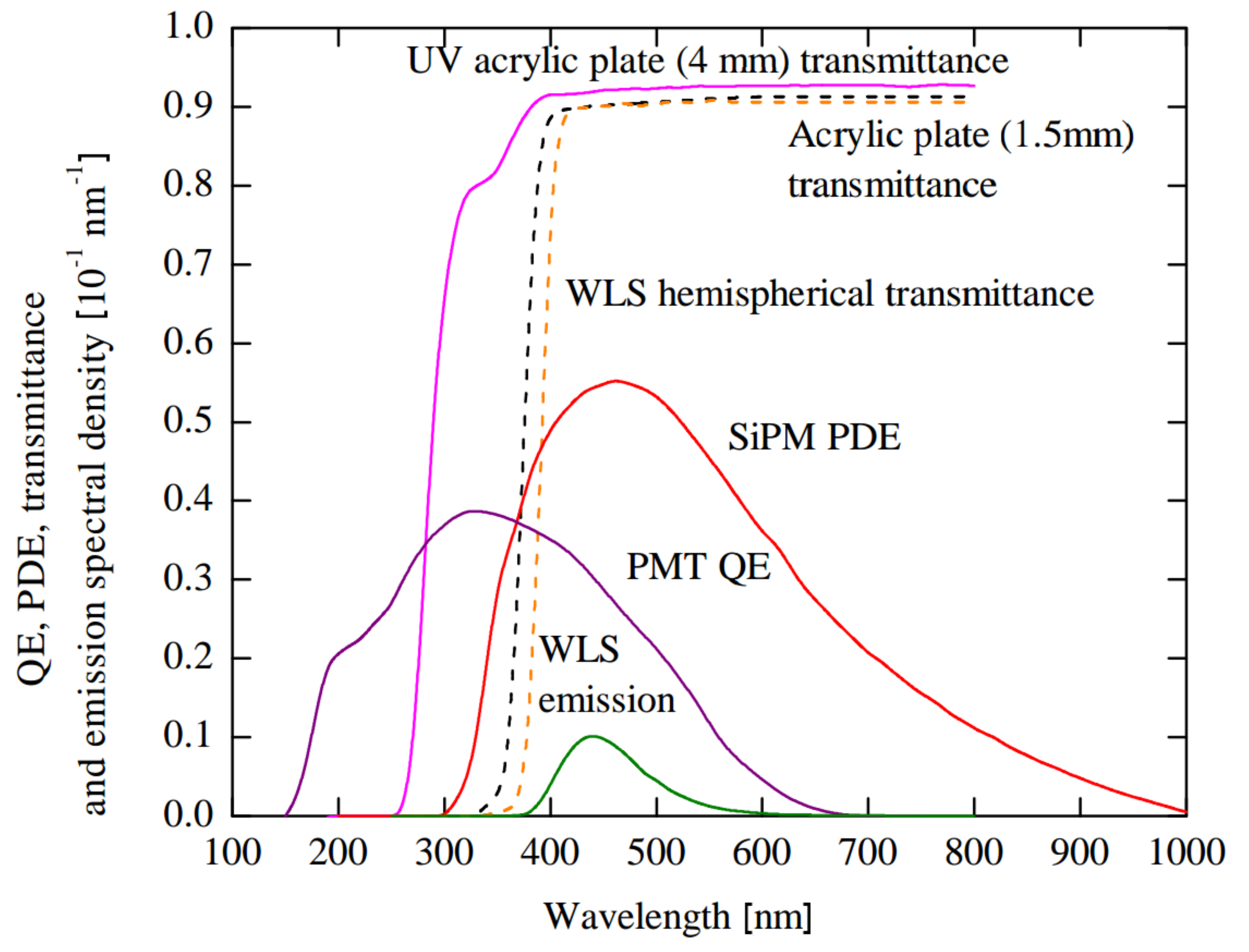}
	\caption{Quantum efficiency (QE) of the PMT R6041-506MOD at 87~K obtained from \cite{Hamamatsu, Lyashenko2014:PMTs} using a temperature dependence derived there, Photon Detection Efficiency (PDE) of the SiPM (MPPC 13360-6050PE \cite{Hamamatsu}) at an 5.6~V overvoltage obtained from \cite{Otte2017} using the PDE voltage dependence, transmittance of the ordinary and UV acrylic plate in front of the SiPM and bare PMT respectively, measured by us, and hemispherical transmittance of the WLS (TPB in polystyrene) \cite{Francini2013}. Also shown is the emission spectrum of the WLS (TPB in polystyrene) \cite{Gehman2013}.}
	\vspace{-10pt}
	\label{fig:spectra}
  	\end{figure}

The voltage applied to the divider may vary from 3 to 22~kV providing the electric field of 0.093-0.68~kV/cm in liquid Ar in the drift region (between the cathode and THGEM0), 0.71-5.2~kV/cm in liquid Ar in the electron emission region (above THGEM0) and 1.1-8.0~kV/cm in gaseous Ar in the EL region (between the liquid surface and THGEM1). All electrodes had the same active area of 10$\times$10~cm$^2$. This geometry was confirmed to have satisfactory field uniformity \cite{Frolov2019:FieldSimulations}.

Two types of devices were used for optical readout in the detector. The first one is four compact 2-inch R6041-506MOD PMTs \cite{Bondar2015:PMTs, Bondar2017:PMTs} placed on every side of EL gap (Fig.~\ref{fig:setup_scheme}). Another one is 5$\times$5 13360-6050PE SiPM matrix \cite{Hamamatsu} located above the anode. Each PMT and SiPM channel is read out and recorded individually at 62.5~MS/s rate for a total waveform of 160~$\mu$s width with 25~$\mu$s pre-trigger duration. Faster digitization at 250~MS/s was also possible in order to have more samples for prompt photoelectron (PE) peak, but due to limited channel number it was possible to record only 4 PMTs and single SiPM channel in that case.

In this work, two versions of the detector were used. In the first one, a wavelength shifter (WLS) film (tetraphenyl butadiene (TPB) in polystyrene matrix) was deposited in front of 3 (out of 4) side PMTs in order to provide sensitivity to ordinary (excimer) VUV emission of Ar. They are further referred to as 3PMT+WLS. The single bare PMT was made sensitive from UV to visible light by replacing ordinary acrylic in front of it to UV acrylic, see Fig.~\ref{fig:spectra}. MC simulations and experimental data showed that SiPM matrix located above THGEM1 anode was not sensitive to cross-talk due to WLS re-emission \cite{NBrS18}. The bare PMT, however, had noticeable cross-talk in the non-VUV from 3PMT+WLS. The parameters of all THGEMs were similar to those of \cite{THGEMPaper13}: dielectric thickness of 0.4~mm, hole pitch of 0.9~mm, hole diameter of 0.5~mm and hole rim of 0.1~mm.

In the second version of the detector, the WLS was removed altogether and replaced by UV acrylic box in order to study Ar non-VUV emission in more detail and "purer" conditions \cite{NBrS18, NBrS20}. Besides that, using results from \cite{Frolov2019:FieldSimulations} the THGEM0 and THGEM1 were changed to those with larger (75\%) optical transparency with the following parameters: dielectric thickness of 0.94~mm, hole pitch of 1.1 mm, hole diameter of 0.5 mm and without hole rims. In addition, we changed R3 resistance on THGEM0 (Fig.~\ref{fig:setup_scheme}) from 4 to 10~M$\Omega$ and replaced field shaping electrodes from those of opaque plates to simple metallic wires. All these modifications increased electron transmission through THGEM0 to 100\% and improved light collection efficiency for both S1 and S2 signals.

To summarize, we analyze and discuss the results in terms of 5 groups of devices: 3PMT+WLS, SiPM matrix(WLS) and bare PMT(WLS) for the first configuration of the detector (with WLS) and 4PMT and SiPM matrix for the second configuration (without WLS). It is important to note that in both cases SiPM matrix was not sensitive to VUV re-emitted by WLS; "SiPM matrix(WLS)" was used to merely distinguish the SiPM data taken in different conditions.

A $^{109}$Cd gamma-ray source was used in the measurements. The emission spectrum of this source includes low-energy (22-25~keV) and high-energy lines, namely the characteristic lines of W(60~keV), which was used as a radionuclide substrate, and the 88~keV line of $^{109}$Cd itself \cite{Bondar2019:Cd_source}. Due to insufficient energy resolution, the 60 and 88 keV lines could not be separated; therefore, their weighted average energy (82~keV \cite{Bondar2019:Cd_source}) was used in the analysis.

The measurements of S2 pulse-shapes were conducted mainly using $^{109}$Cd gamma-ray source in several runs and with one run using $^{238}$Pu alpha-particle source installed at the center of the cathode. The latter data are not presented in this work, but preliminary analysis confirms the $^{109}$Cd results. Since the S1 signal from radioactive sources was too weak, the S2 signal was used for the trigger. Namely, a threshold trigger was used on the sum of 3PMT+WLS channels, amplified with 200~ns shaping, in the first detector version. In the second version, the sum of 4PMT channels was used instead with an additional 2~$\mu$s shaping amplifier. Such amplification was used because the number of photoelectrons was quite small without WLS and in order to distinguish events from single PE noise one had to effectively integrate the S2 signal over 2 $\mu$s time window.

The main results on S2 pulse-shapes were obtained at 62.5~MS/s sampling rate and for $^{109}$Cd gamma source in the range of reduced electric field in EL gap from 3.4 to 8.5~Td (8 to 20~kV of high voltage). No difference between 62.5 and 250~MS/s S2 pulse-shapes results was observed. In other words, 250 MS/s data were used only for confirmation of the main data set. In the same manner, $^{238}$Pu alpha-particle source as well was only for confirmation of independence of the results on S2 amplitude and localization of primary ionization. Thus, in the next sections we present and discuss only 62.5~MS/s and $^{109}$Cd data obtained in a single experimental run for detector configuration with WLS and compiled data recorded at three separate runs for the configuration without WLS.
	
	\section{Analysis algorithm}
	
	\begin{figure}[!t]
	\centering
	\includegraphics[width=\linewidth]{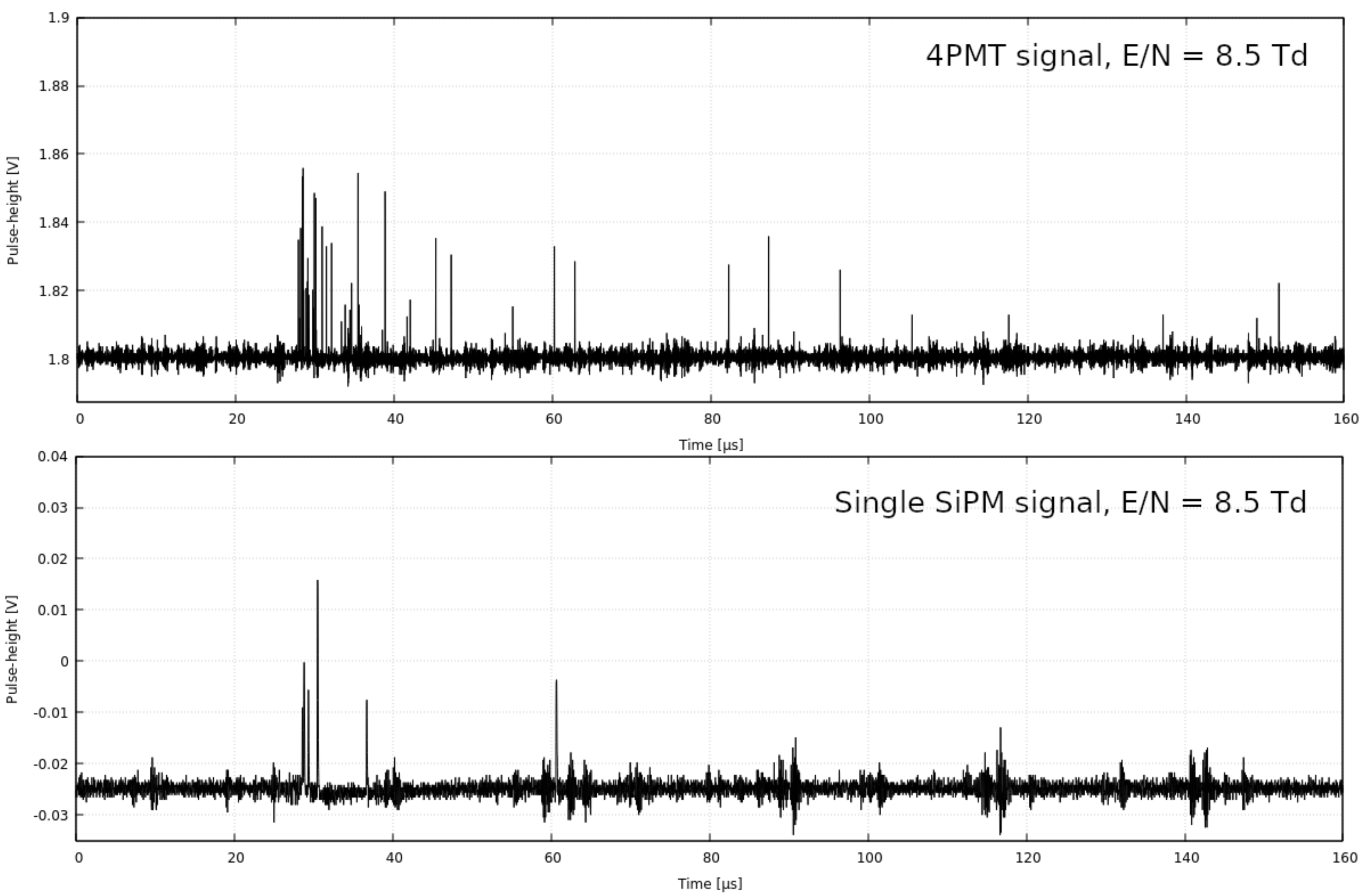}
	\caption{Examples of raw signals for sum of four PMTs (top) and central SiPM channel (bottom) at the maximal reduced electric field in the EL gap, of 8.5~Td, for the detector configuration without WLS.}
	\label{fig:raw_signals}
  	\end{figure}

	\begin{figure}[t]
	\centering
	\includegraphics[width=\linewidth]{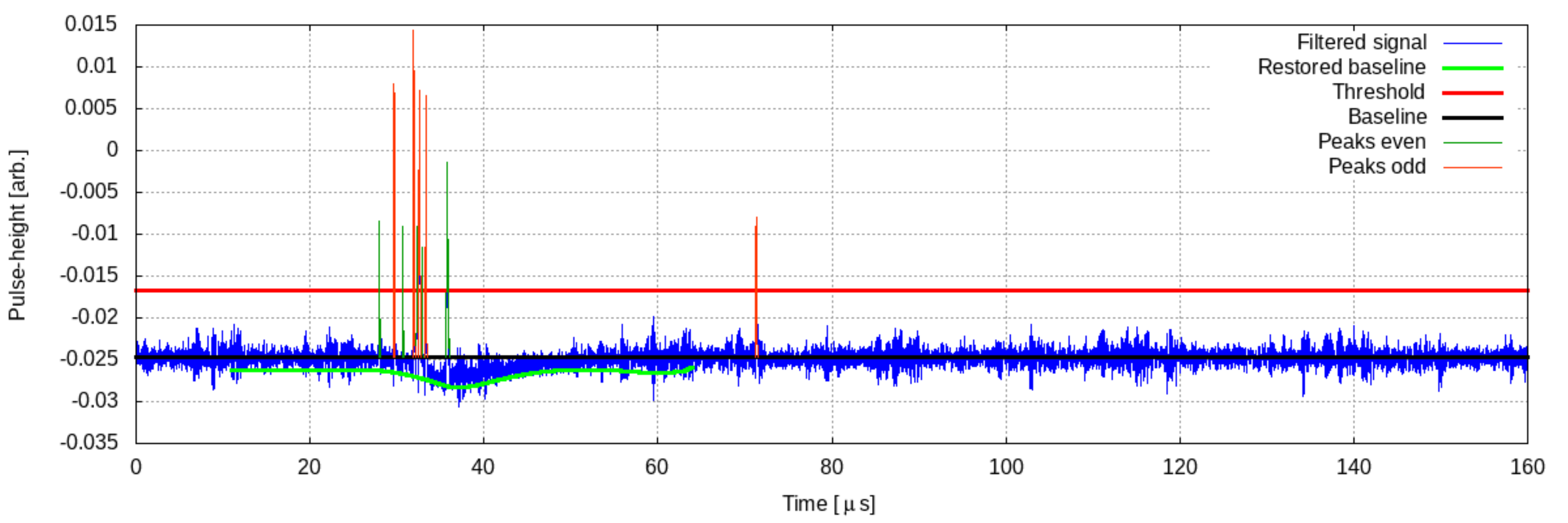}
	\caption{Example of filtered SiPM signal with restored baseline. It is the event at the maximal reduced electric field in the EL gap, of 8.5~Td, for the detector configuration without WLS.} 
	\vspace{-10pt}
	\label{fig:signal_processed}
  	\end{figure}

	The first step of the analysis was to extract PE peaks from the raw waveforms. The examples of the signals are shown in Fig.~\ref{fig:raw_signals}. Simple threshold algorithm was used for PE peak selection. In some cases, however, we also had to apply additional Savitzky-Golay filtering. Moreover, some signals had significant negative afterpulses as demonstrated in Fig.~\ref{fig:signal_processed}. To account for that, we used baseline restoration algorithm \cite{Ryan1988:bkg_algorithm} implemented in ROOT package \cite{CERN_ROOT}. The result of this algorithm is showed by green line in Fig.~\ref{fig:signal_processed}. For faster computations, this algorithm was applied only near S2 region. Systematical shift of restored baseline from the real baseline determined from the pre-trigger region (black line) was accounted for. After that, calibrations of each channel, i.e. calculations of PE peak average area, were conducted using pre-trigger region of the signal.
	
	\begin{figure}[t]
	\centering
	\includegraphics[width=\linewidth]{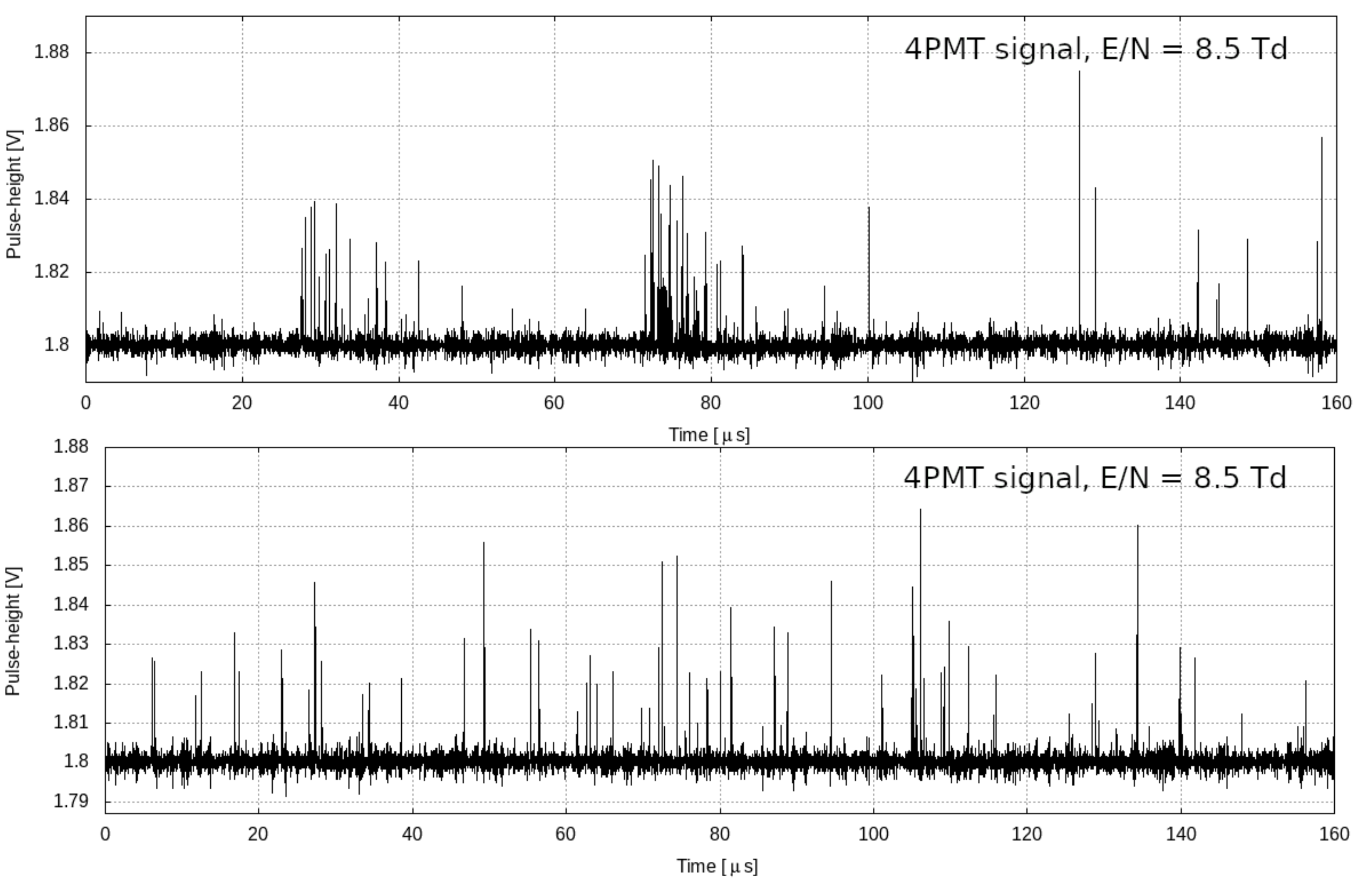}
	\caption{Examples of events which were rejected by the analysis algorithm. The events are for the detector configuration without WLS, at the maximal reduced electric field in the EL gap, of 8.5~Td.}
	\label{fig:signals_rejected}
  	\end{figure}

	\begin{figure}[t]
	\centering
	\includegraphics[width=0.7\linewidth]{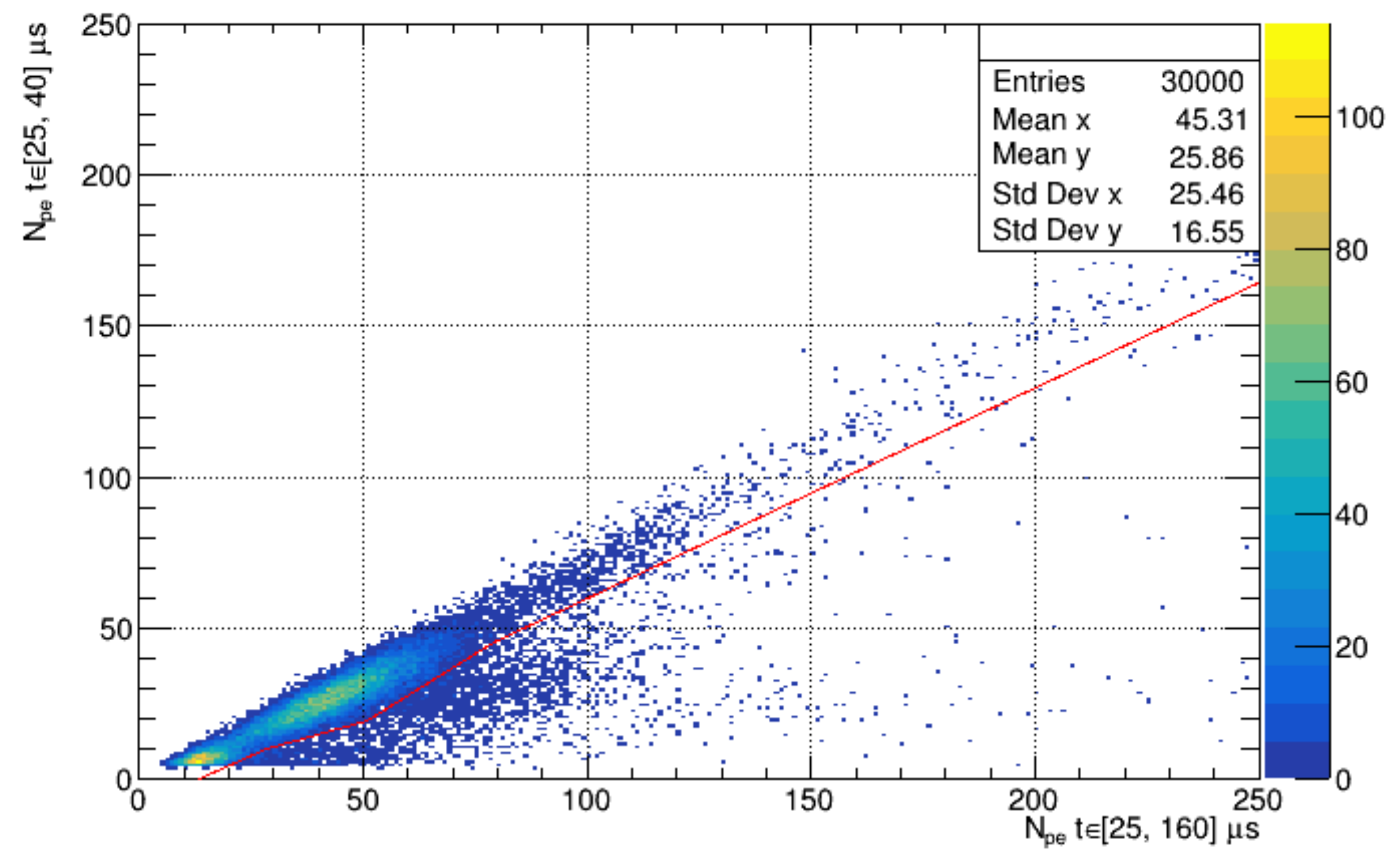}
	\caption{2D plot of 4PMT amplitudes (expressed in  photoelectron number $N_{pe}$) used for rejection of bad events (see text for details). The area to the right of red line corresponds to the events shown in Fig.~\ref{fig:signals_rejected}. The cluster at around (45, 25) corresponds to the 82~keV $^{109}$Cd line. This plot is for the detector configuration without WLS, at the maximal reduced electric field in the EL gap, of 8.5~Td.} 
	\vspace{-10pt}
	\label{fig:2D_Npe_plot}
  	\end{figure}

	After extracting PE peaks and calibrating we proceeded to higher-level analysis, namely selection of events and producing S2 pulse-shapes. At first, events with two superimposed S2 signals or events consisting of PE "trains" as shown in Fig.~\ref{fig:signals_rejected} were rejected. This rejection was based on 2D plot of number of PMTs photoelectrons in small time window around S2 signal (25-40~$\mu$s) versus number of PE in full time range (25-160~$\mu$s) for each event. The example of this plot as well as region corresponding to the rejected events is shown in Fig.~\ref{fig:2D_Npe_plot}. Besides visual confirmation that the area on the right corresponds to abnormal events, we compared this plot for data taken with and without Cd source at the same measurement and analysis conditions. In both cases rejection area contains these events, leading us to the conclusion that they do not correspond to the real Cd signal. 

	\begin{figure}[t]
	\centering
	\includegraphics[width=0.7\linewidth]{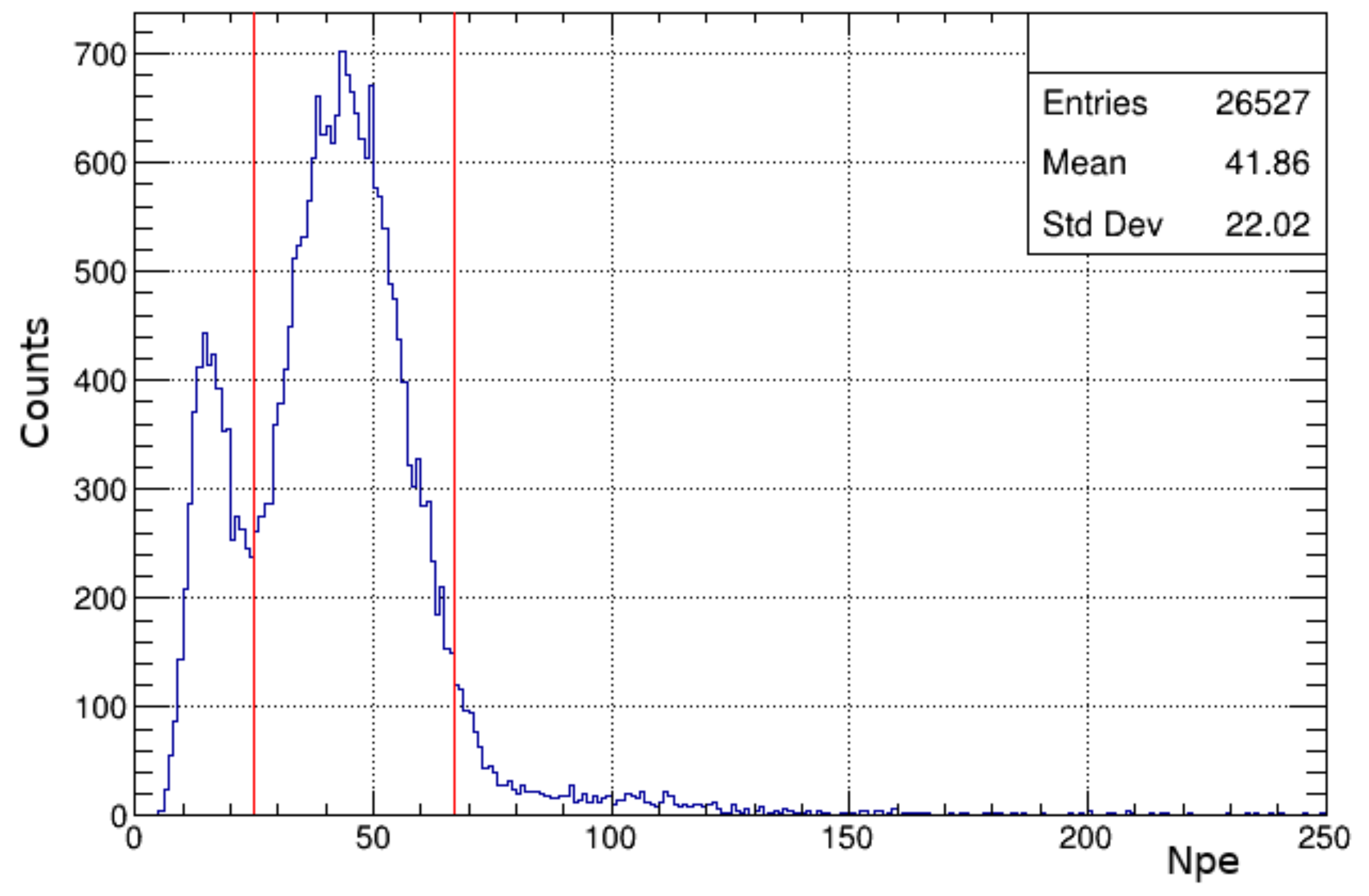}
	\caption{Amplitude spectrum in the two-phase detector for $^{109}$Cd gamma-ray source, corresponding to the projection of Fig.~ \ref{fig:2D_Npe_plot} on x-axis. This distribution was plotted after cuts shown there were applied. Lines show cuts which were used for the event selection.} 
	\vspace{-10pt}
	\label{fig:Cd_spectrum}
  	\end{figure}

	Finally, only events corresponding to 82~keV peak of the $^{109}$Cd gamma-ray source are selected using histogram of PE number for all PMTs, its example shown in Fig.~\ref{fig:Cd_spectrum}. Bad events are already rejected in this histogram. As can be seen, there is still some background present which is primarily caused by cosmic muons. Changing selection criteria (e.g. selecting not the whole 82~keV peak, but only one of its slopes) showed no significant effect on S2 pulse-shapes. After all described cuts, we build peak time histograms weighted with the number of photoelectrons in that peak which was calculated as area of the peak divided by area of a single PE obtained in the calibration. These peak time histograms are further referred to as S2 pulse-shapes. Also, all following pulse-shape results will have baseline determined from region before the trigger already subtracted.

	\section{Results}
	\subsection{Data without WLS}\label{results_no_WLS}
	
\begin{figure}[t]
	\centering
	\includegraphics[width=\linewidth]{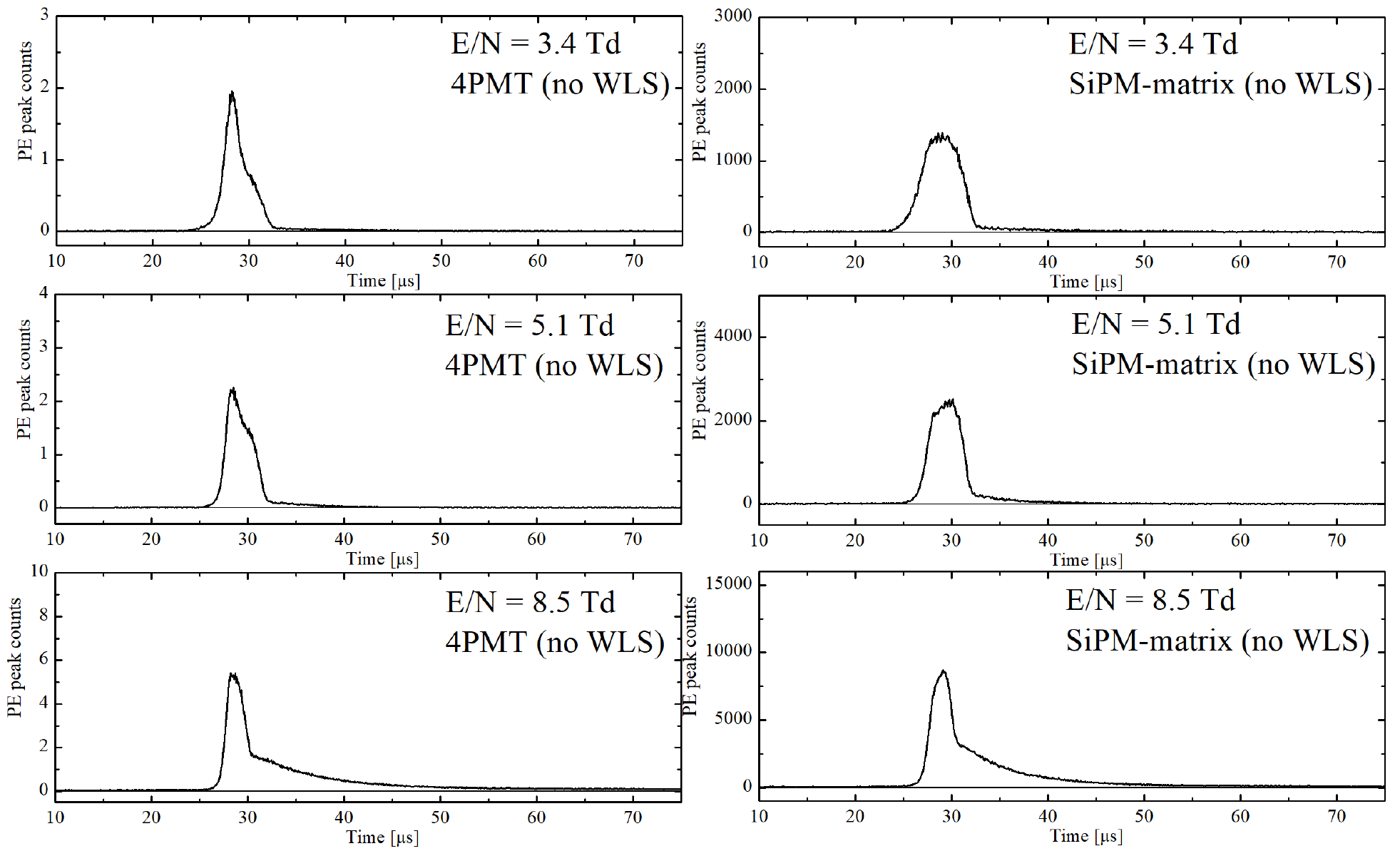}
	\caption{S2 pulse-shapes in two-phase Ar detector in configuration without WLS, for the 4PMT (left) and SiPM-matrix (right) readout at different reduced electric fields.} 
	\vspace{-10pt}
	\label{fig:origin_no_WLS}
  	\end{figure}

	First we present the data on the S2 pulse-shapes for the detector configuration without WLS, i.e. when PMTs and SiPM matrix were not sensitive to the VUV and consequently to the 3.1~$\mu$s excimer component, see Fig.~\ref{fig:origin_no_WLS}. As expected, at low fields, below the threshold of ordinary EL in the VUV (4~Td), only the fast component produced by neutral bremsstrahlung (NBrS) mechanism \cite{NBrS18, NBrS20} is observed. 
	
	There is distortion of its shape compared to the expected rectangular-like shape (see section~\ref{PulseShapeOverview}) due to the fact that the trigger is taken from the S2 signal itself. Indeed, the pulse shape is distorted because the trigger is fired by the first photoelectron (PE) signal matching trigger selection criteria, defined by particular threshold, amplifier shaping time and specific pulse shape. It is obvious that in this case there is always a PE peak at the trigger position, which is smoothed out at high PE statistics and stands out at low PE statistics.

	\begin{figure}[t]
	\centering
	\includegraphics[width=\linewidth]{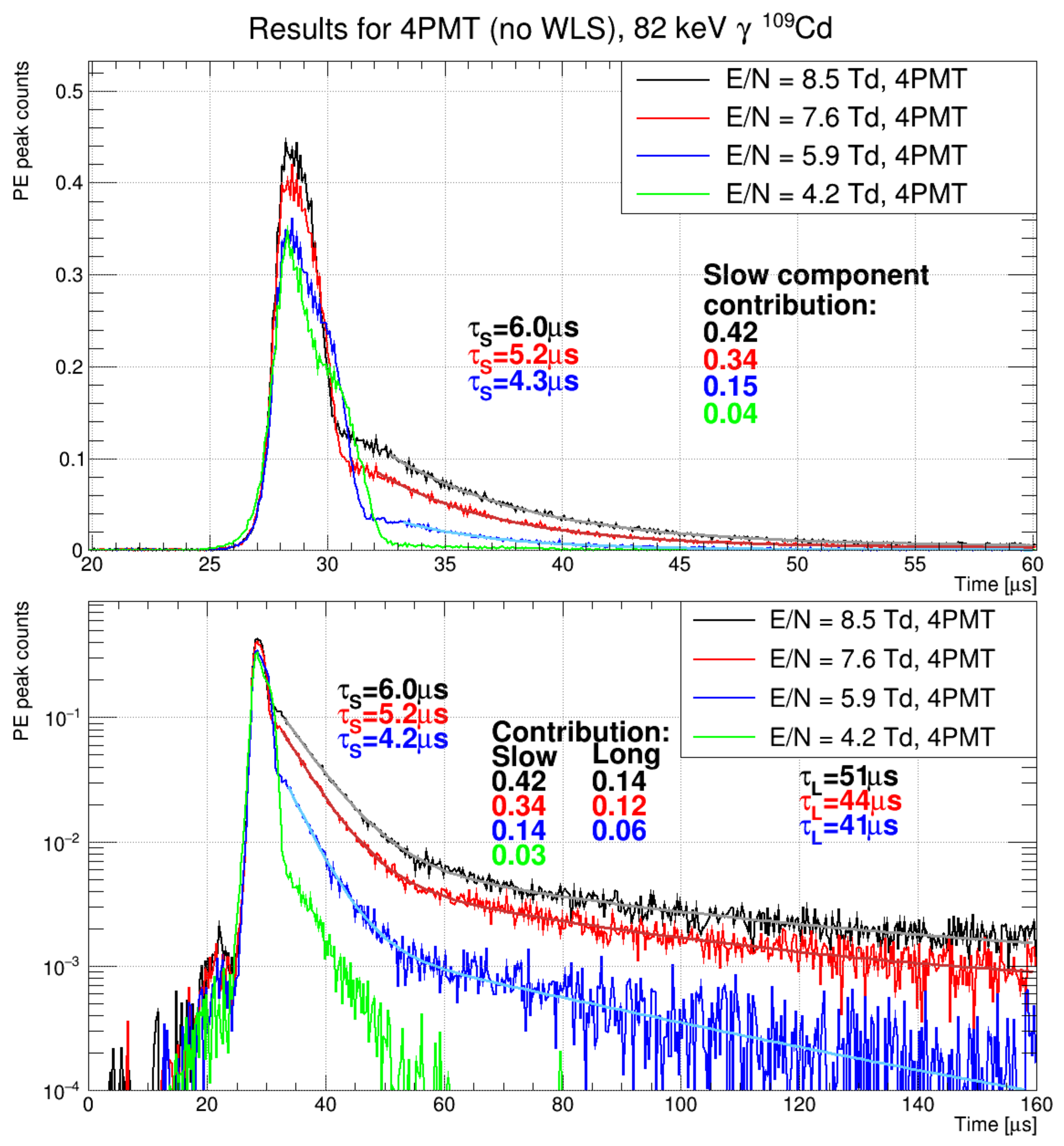}
	\caption{S2 pulse-shape in two-phase Ar detector in configuration without WLS for the 4PMT readout in linear (top) and logarithmic (bottom) scale. The lines show the fit of slow and long components. The pulse-shapes are normalized by area of the fast components. See text for details on fitting procedure and parameter definitions.}
	\label{fig:fit_definition}
  	\end{figure}

	On the other hand, the unusual slow component was unexpectedly observed at higher electric fields for both 4PMT and SiPM-matrix readout, with time constant varying from 4.0 to 5.6~$\mu$s, and its contribution increasing with the electric field.
	 In addition, another component was observed with even larger time constant, of about 40~$\mu$s, on logarithmic scale, see Fig.~\ref{fig:fit_definition}. It is further referred to as long component. This figure also demonstrates the fitting procedure used to quantify the slow and long components for all electric fields and detector configurations. Since the origin of these components is unknown, a simple fit formula reflecting their exponential behaviour was used:
	\begin{equation} \label{eq:1}
	f(t)\,=\,y_0\,+\,A_{S,L}\ast\textrm{exp}\left(-\frac{t - t_0}{{\tau}_{S,L}}\right),\:t\,>\,t_0
	\end{equation}
where $y_0$ is the baseline, $A_{S,L}$ are the amplitudes of the slow and long components, ${\tau}_{S,L}$ are their decay times (time constants), and $t_0$ defines the beginning of the fit region. The last parameter was fixed by hand. The contribution of the slow+long components was defined as the ratio of the S2 pulse-shape area right after the fast component (t~>~30.5~$\mu$s for 8.5~Td in Fig.~\ref{fig:fit_definition}.) to the total area. The long component contribution was defined using the area of its fit.

	It should be noted that the separation of fast and slow components based on simple time threshold may result in a certain systematic error. Nevertheless, it suits the purposes of demonstrating the characteristic properties of these components.  

\begin{figure}[t]
	\centering
	\includegraphics[width=0.7\linewidth]{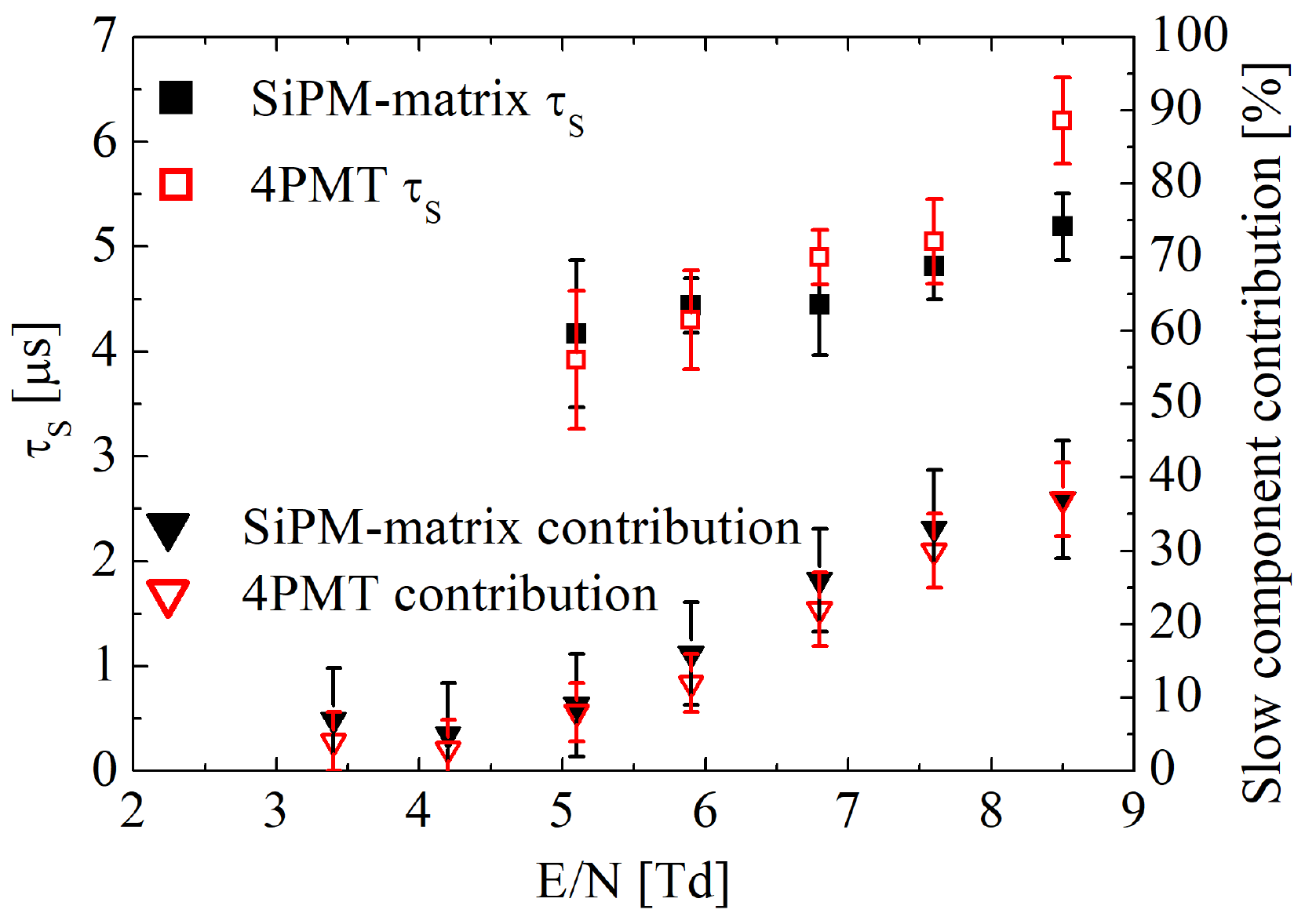}
	\caption{Time constant $\tau_{S}$ and contribution of slow component vs electric field in EL gap for the detector configuration without WLS.}
	\vspace{-10pt}
	\label{fig:slow_comp_no_WLS}
	\end{figure}

\begin{figure}[t]
	\centering
	\includegraphics[width=0.7\linewidth]{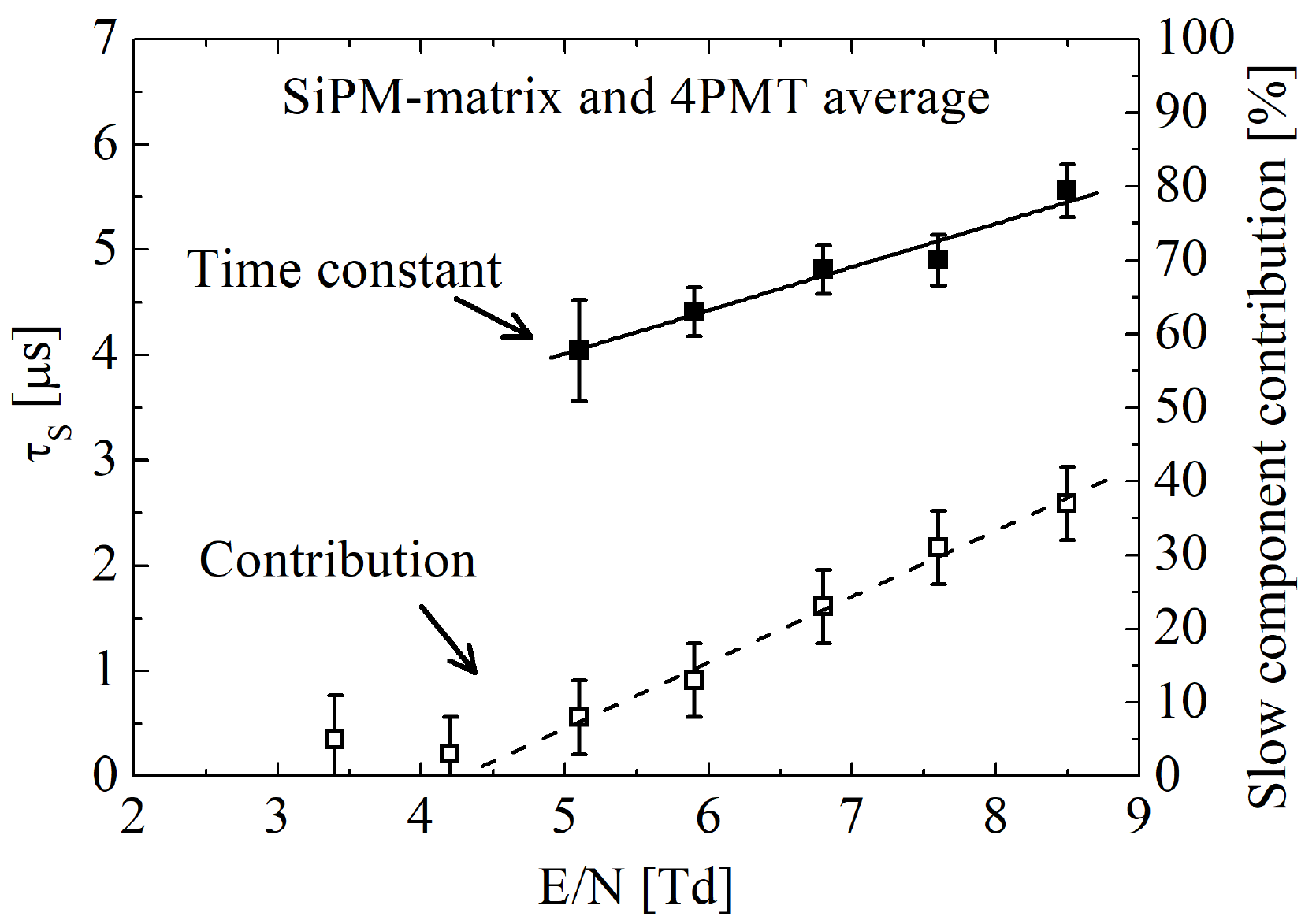}
	\caption{Time constant $\tau_{S}$ and contribution of slow component taken as average between the 4PMT and SiPM-matrix data (Fig. \ref{fig:slow_comp_no_WLS}) for the detector configuration without WLS.}
	\vspace{-10pt}
	\label{fig:slow_comp_no_WLS_avg}
  	\end{figure}

	The dependence of the slow component time constant and contribution on the electric field is shown in Fig.~\ref{fig:slow_comp_no_WLS} for both 4PMT and SiPM-matrix readout. Since the data for both readout configurations are in good agreement, one can take their average which is shown in Fig.~\ref{fig:slow_comp_no_WLS_avg}. Perhaps the most important result of this work is that both time constant and contribution of the slow component increase with the increase of the electric field. It can't be explained by any standard mechanism discussed in section~\ref{PulseShapeOverview}.


	The errors shown in the Fig.~\ref{fig:slow_comp_no_WLS},~\ref{fig:slow_comp_no_WLS_avg} and all figures in the next sections are both statistical and systematic, the former being much less than the latter. The systematic errors have three main sources: the error due to the histogram binning, the error due to fitting procedure, and the error due to analysis algorithms, i.e. event selection. The last error was estimated by comparing the results for events selected from the 82~keV peak as shown in Fig.~\ref{fig:Cd_spectrum} with the events belonging to only the left slope of the peak. Each of the systematic errors was estimated for both 4PMT and SiPM-matrix readout for each electric field. They turned out to be of similar magnitude.

	In addition, the small tail of "slow component" sometimes seen at the reduced fields below 4.5~Td is attributed to the background of non $^{109}$Cd events mistakenly passing the event selection. 
	This conclusion mas made by comparing different measurement runs and selection criteria, which sometimes showed non-exponential behavior of the "slow component" at these fields. 

In principle, there are two general mechanisms through which the unusual slow component can be produced: either due to some unknown scintillation mechanism, which has its time constant and intensity dependent on the electric field, or this slow component is present in the charge signal itself.
	In order to discern the mechanism of the unusual slow component the 3PMT+WLS data  should be considered.
	
	\subsection{Data with WLS}\label{results_with_WLS}

\begin{figure}[t]
	\centering
	\includegraphics[width=\linewidth]{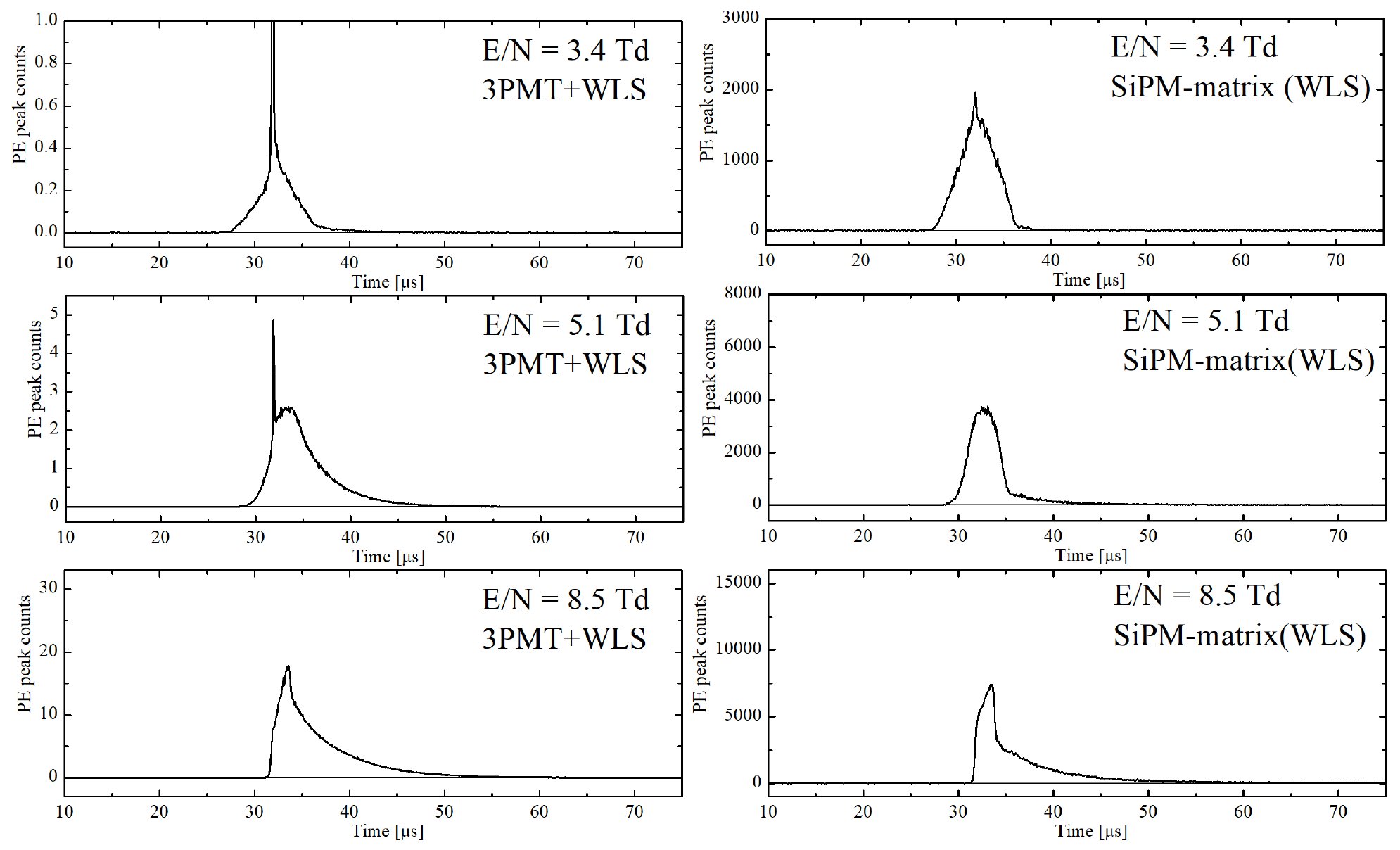}
	\caption{S2 pulse-shape in two-phase Ar detector in configuration with WLS, for the 3PMT+WLS (left) and SiPM-matrix(WLS) (right) readout at different electric fields.} 
	\vspace{-10pt}
	\label{fig:origin_with_WLS}
  	\end{figure}

	The examples of S2 pulse-shapes obtained in the detector configuration with WLS are shown in Fig.~\ref{fig:origin_with_WLS}. As in the case without WLS, only fast component is observed below the 4~Td threshold of ordinary EL. The sharp peaks on 3PMT+WLS pulse-shapes are artifact caused by the trigger which was taken from this signal itself (see section~\ref{results_no_WLS}).

	\begin{figure}[t]
	\centering
	\includegraphics[width=0.7\linewidth]{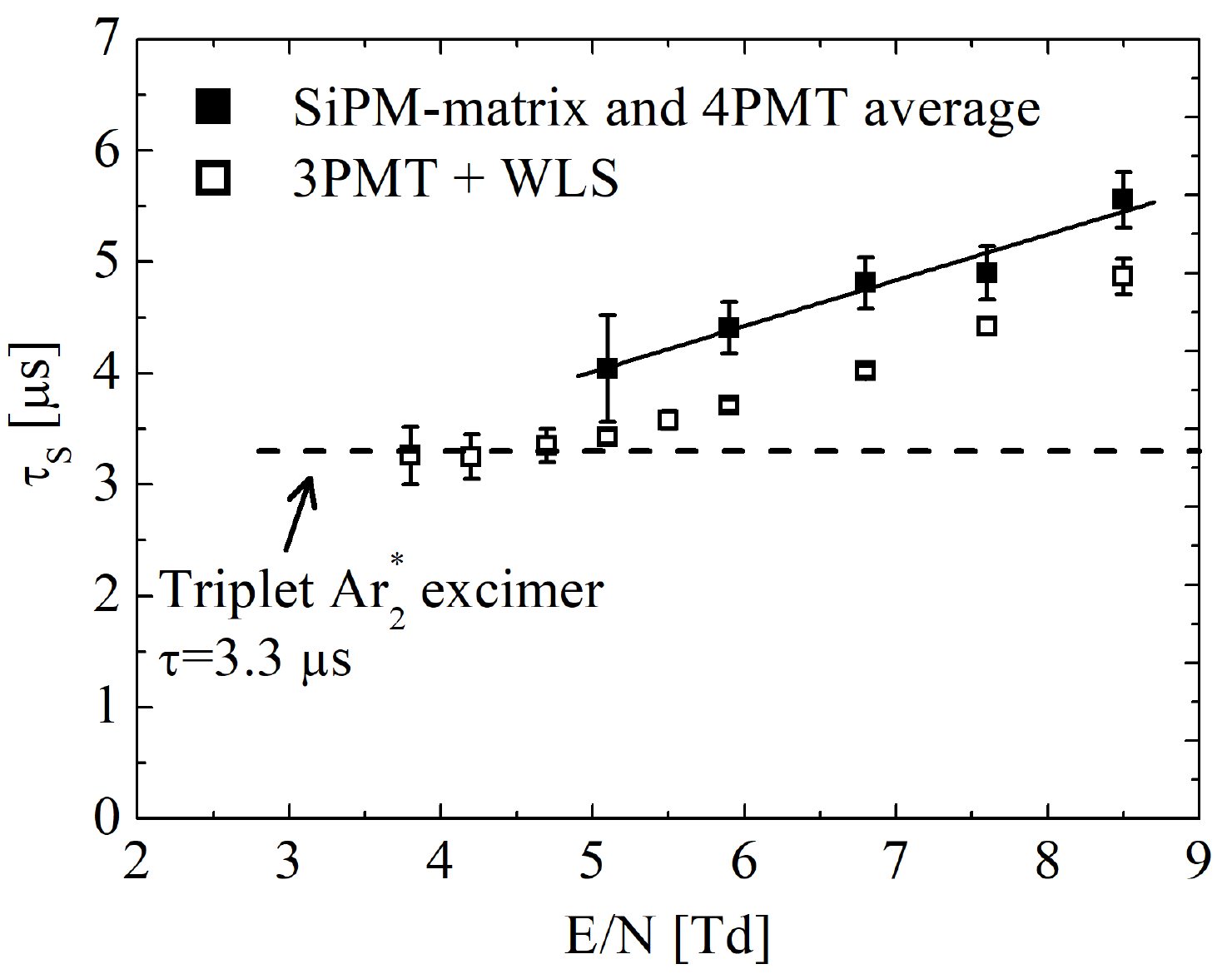}
	\caption{Comparison of slow component time constant ($\tau_{S}$) dependence on the electric field for the data without WLS (average of 4PMT and SiPM-matrix data) and with WLS (3PMT+WLS data).} 
	\vspace{-10pt}
	\label{fig:slow_comp_tau_with_WLS}
  	\end{figure}

	\begin{figure}[t]
	\centering
	\includegraphics[width=0.7\linewidth]{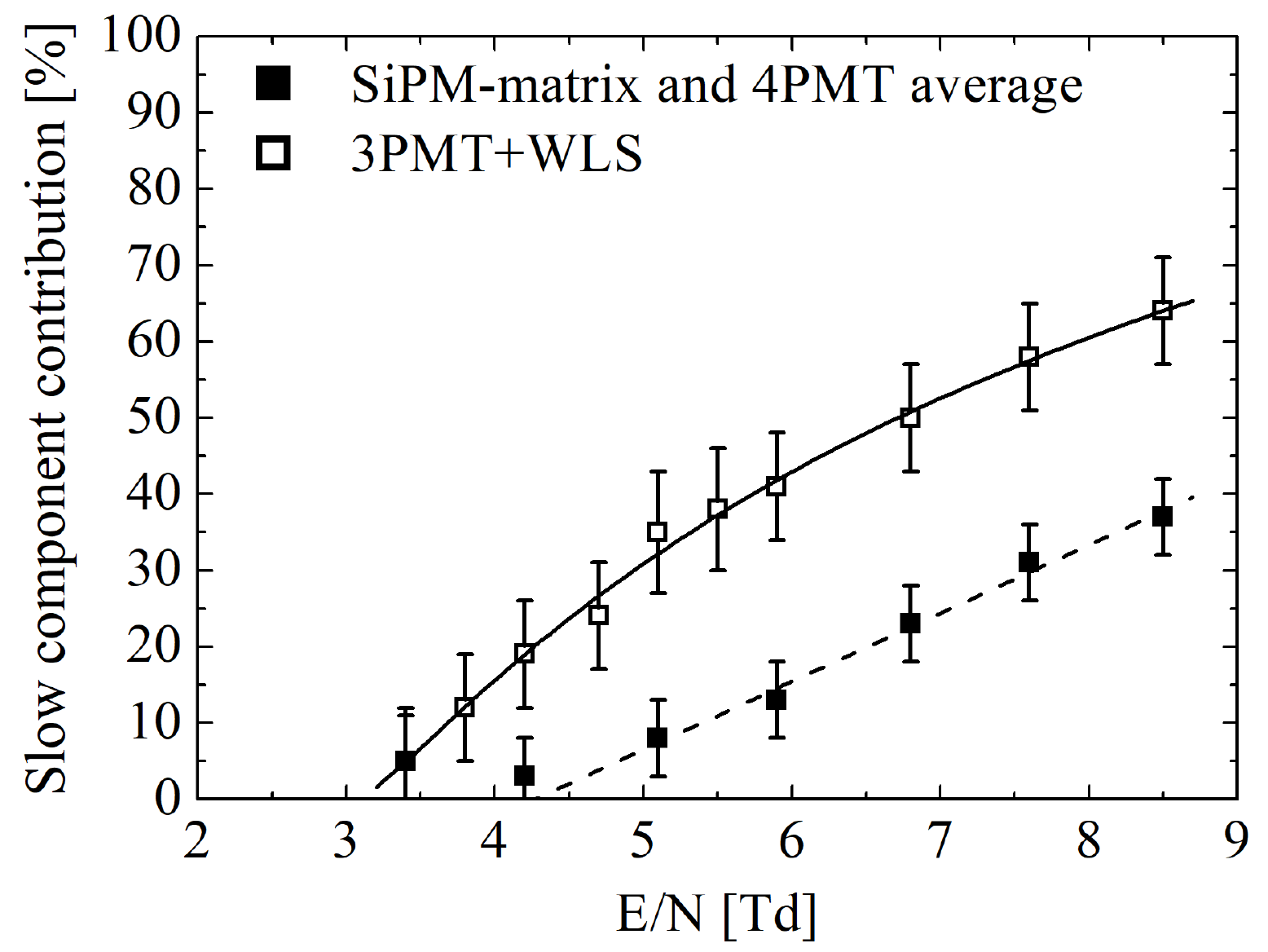}
	\caption{Comparison of slow component contribution dependence on the electric field for the data without WLS (average of 4PMT and SiPM-matrix data) and with WLS (3PMT+WLS data).} 
	\vspace{-10pt}
	\label{fig:slow_comp_contr_with_WLS}
  	\end{figure}
	
	Another important observation is that SiPM-matrix(WLS) pulse-shapes shown in Fig.~\ref{fig:origin_with_WLS} are almost identical to those of SiPM matrix in the detector configuration without WLS shown in Fig.~\ref{fig:origin_no_WLS}. The comparison of their time constant and contribution values shows an excellent agreement. This confirms the results of the previous experiments and simulations showing that the SiPM matrix had no cross-talk due to WLS re-emission \cite{NBrS18}.

On the other hand, the 3PMT+WLS pulse-shapes are clearly different from those of 4PMT and are close to what would be expected from the triplet excimer emission mechanism \cite{Agnes2018}. In particular, there is no distinct bend separating fast and slow component compared to that seen in SiPM-matrix signals, apparently due to the larger contribution induced by the triplet excimer emision. 

The 3PMT+WLS pulse-shapes were fitted according to the same procedure as described in the previous section~\ref{results_no_WLS}. The results for the time constant are presented in Fig.~\ref{fig:slow_comp_tau_with_WLS}. At the fields in the range of 4.0-5.1~Td, where the contribution of the unusual slow component observed in the data without WLS is close to zero (Fig.~\ref{fig:slow_comp_no_WLS_avg}), one can see that the 3PMT+WLS ${\tau}_{S}$ is independent of the field and amounts to about 3.3$\pm$0.1~$\mu$s which agrees with the known value of the triplet excimer decay time of 3.1 $\mu$s \cite{Buzulutskov17:ELReview}. At higher electric fields, however, the 3PMT+WLS time constant increases with the field, at the same time being smaller than the slow component time constant of the data without WLS. Such a characteristic behaviour of the time constant is easiest to explain assuming that the unusual slow component is present in the charge signal itself, meaning it is present regardless of the optical readout configuration. 

The second thing to consider is the dependence of the 3PMT+WLS slow component contribution on the electric field, see Fig.~\ref{fig:slow_comp_contr_with_WLS}. Since the VUV emission is dominant over neutral bremsstrahlung mechanism by at least an order of magnitude \cite{NBrS18, NBrS20}, the triplet excimer component contribution should rapidly increase near the 4~Td threshold and reach a plateau. In contrast, Fig. \ref{fig:slow_comp_contr_with_WLS} shows no plateau for the 3PMT+WLS data, indicating that the contribution continues to grow, obviously due to the unusual slow component contribution.

	Finally, the charge signal hypothesis can naturally explain that SiPM matrix and 4PMT have the same pulse-shapes despite having different spectral characteristics, see Fig.~\ref{fig:spectra}. 
	
	To conclude, we propose that the slow component may be produced in charge signal if electrons are somehow delayed or trapped during their drift in the EL gap. This is possible via formation of metastable negative argon ions (Ar$^-$). There are two possible candidates for such states: the Feshbach resonances Ar$^{-}(3p^{5}4s^{2}\:^{2}P_{1/2,3/2})$ \cite{Kurokawa2011:Ar_Feshbach} or Ar$^{-}(3p^{5}4s4p\:^{4}S)$ state with $\sim$300~ns lifetime \cite{Bae1985:Ar_ion_state, Ben-Itzhak1988:Ar_ion_state}. In this case, the fast component corresponds to the electrons which drift through the EL gap without being trapped. The advantage of this approach is that it can explain the increase of slow component contribution with the electric field: the formation of metastable ions has energy threshold and the higher the electric field is, the larger the probability of electron having enough energy to become trapped.

	The proposed mechanism of electron trapping predicts the dependence of the slow component contribution on the EL gap thickness: larger thickness means higher probability that electron will form metastable ion during its drift. In order to verify our hypothesis one has to study the dependence of S2 pulse-shapes on the EL gap thickness.
		
	\section{Conclusion}

	We have for the first time systematically studied S2 pulse-shapes in two phase Ar detector in a wide range of reduced electric fields, from 3.4 to 8.5~Td, with different readout configurations, using SiPM matrix and PMTs, and in different spectral ranges, in the VUV and visible range. The unusual slow and long components have been observed in the visible range with time constants of $\sim$5~$\mu$s and $\sim$40~$\mu$s respectively. The unusual character of the slow component was that its time constant and contribution increased with the electric field from 4.0~$\mu$s and $\sim$8\% at 5.1~Td to 5.5~$\mu$s and $\sim$37\% at 8.5~Td.
	
	Such behaviour of the slow component can not be explained by any known mechanism. We propose that this component is present in the charge signal itself, presumably due to electron trapping on metastable negative argon ions (Ar$^-$). In order to verify this hypothesis we plan to study the dependence of S2 pulse-shapes on the EL gap thickness. It would be another convincing evidence to directly observe the slow component in the charge signal, which is included in our future plans as well.
	
	Regarding the long component, its properties and nature are poorly understood at the moment. Further investigation is required.   
	
	The results obtained can have practical applications in dark matter search experiments based on two-phase Ar, in particular in the DarkSide experiment.

	\acknowledgments
This work was supported in part by Russian Science Foundation (project no. 20-12-00008). The work was done within the R\&D program for the DarkSide-20k experiment.
	\bibliography{Manuscript} 
\end{document}